\documentclass[twocolumn,showpacs,preprintnumbers,amsmath,amssymb,superscriptaddress]{revtex4}
\usepackage{longtable}
\usepackage{graphicx}

\usepackage{dcolumn}
\usepackage{bm}

\begin{document}

\title {Atmospheric Charged $K/\pi$ Ratio and Measurement of Muon Annual Modulation with a Liquid Scintillation Detector at Soudan}
\newcommand{\usd}{Department of Physics, The University of South Dakota, Vermillion, South Dakota 57069}
\newcommand{\mmc}{Natural Sciences Division, Mount Marty College, Yankton, South Dakota 57078, USA}

\newcommand{\dmm}{Dongming.Mei@usd.edu}

\author{C. Zhang} \affiliation{\usd} \affiliation{\mmc}
\author{D.-M. Mei} \altaffiliation[Corresponding Author: ]{\dmm}  \affiliation{\usd}

\begin{abstract}
We report a measurement of muon annual modulation in a 12-liter liquid scintillation detector with a live-time of more than 4 years at the Soudan Underground Laboratory. Muon minimum ionization in the detector is identified by its observed pulse shape and large energy deposition. The measured muon rate in the detector is 28.69$\pm$2.09 muons per day with a modulation amplitude of 2.66$\pm$ 1.0\% and a phase at Jul 22 $\pm$ 36.2 days. This annual modulation is correlated with the variation of the effective atmospheric temperature in the stratosphere. The correlation coefficient, $\alpha_{T}$, is determined to be $0.898 \pm 0.025$. This can be interpreted as a measurement of the atmospheric charged kaon to pion ($K$/$\pi$) ratio of 0.094$^{+0.044}_{-0.061}$ for $E_{p} > $ 7 TeV, consistent with the measurement from the MINOS far detector. To further constrain the value of $K$/$\pi$ ratio, a Geant4 simulation of the primary cosmic-ray protons with energy up to 100 TeV is implemented to study the correlation of $K$/$\pi$ ratio and  the muon annual modulation for muon energy greater than 0.5 TeV. We find out that a charged $K$/$\pi$ ratio of 0.1598, greater than the upper bound (0.138) from this work at the production point 30 km above the Earth surface in the stratosphere cannot induce muon annual modulation at the depth of Soudan. 

\end{abstract}

\maketitle

\section{Introduction}
Energetic muons produced by the decay processes of pions and kaons in the stratosphere can reach 
deep underground. Numerous underground detectors discovered the annual modulation of 
muon rates \cite{barret, sherman, torino, hobart, baksan, macro, amanda, borexino, lvd, icecube, minosfar, dmice, doub}. This annual modulation is believed to be correlated with a slow temperature variation over seasons in the stratosphere where those muons are produced. 
Therefore, the muon flux underground has a dependency on the effective temperature of the stratosphere. An increase in the effective
temperature in the stratosphere results in a lower density profile, which decreases the probability of pions and kaons interacting with the atmospheric particles. Consequently, more pions and kaons undergo decays, which increases the numbers of  muons observed in a detector deep underground. Such a correlation factor, named $\alpha_{T}$, between the measured muon flux modulation underground and the effective temperature variation in the stratosphere was studied by several experiments
such as AMANDA~\cite{amanda}, Borexino~\cite{borexino}, MACRO~\cite{macro} and MINOS~\cite{minosfar}. Since kaons and pions in the primary hadronic interactions of cosmic rays in the stratosphere contribute
differently to $\alpha_{T}$ due to the different masses and lifetimes,
it was suggested that measuring the correlation factor, $\alpha_{T}$, between the annual modulation observed in an underground detector
and the temperature variation in the stratosphere can provide the information on the atmospheric charged kaon/pion ($K$/$\pi$)
 ratio~\cite{gra}. This is particularly interesting for an underground site where the observed muons from primary 
cosmic rays can have energies greater than 7 TeV, since the Large Hadron Collider~\cite{lhc} only provides a collision energy
of $\sim$7 TeV. Therefore, the phenomenon of the correlation between the muon annual modulation 
in an underground detector and the temperature variation in the stratosphere 
deserves more experimental and theoretical investigations to understand 
 the expected local behavior of the atmospheric temperature effect and the difference in the muon
flux over a long period of time. Such a strong correlation over a long period of time indicates a stable 
atmospheric charged $K$/$\pi$ ratio, which could shed light on the energies of the primary cosmic rays, and opens a new window for high energy cosmic ray astronomy.    
\par
To monitor the long term flux variation at a deep underground site,
a liquid scintillation detector has been deployed at the Soudan Underground Laboratory (2100 m.w.e)
and run there for over 4 years.
It consists of a meter long and 5 inches in diameter aluminum tube, filled with
12 liters EJ-301 liquid scintillator. Two 5-inch Hamamatsu PMTs (R4144)
are attached to both ends of the tube through Pyrex windows to collect
the scintillation light. Detailed calibration procedures and 
techniques are discussed in Ref. \cite{czDet, czNeu}. 

In a separate paper, we have reported the observation of annual modulation induced by $\gamma$ rays from ($\alpha, \gamma$) reactions at the Soudan Underground Laboratory~\cite{chao}. The energy of the observed $\gamma$ rays is reported in the range of 4 - 10 MeV induced by $\alpha$ particles from radon decays. The amplitude of modulation is found to be 26.5\%, which has been proven to be in correlation with the variation of radon concentration at the Soudan Underground Laboratory. 

In this paper, the variation of the muon rate underground correlating with the
modulation of atmospheric temperature is studied with both experimental data over 4 years and the Monte Carlo simulation
of the primary protons with energies up to 100 TeV. The minimum energy of muons required to reach the depth of 2100 m.w.e. at Soudan Underground Laboratory is above 0.5 TeV. Though the detector is small, the run period is long (4 years) that warrants a meaningful physical result to be reported in this article. 

\section{The variation of muon rate at the Soudan Underground Laboratory}
\begin{figure}
\includegraphics[width=0.45\textwidth]{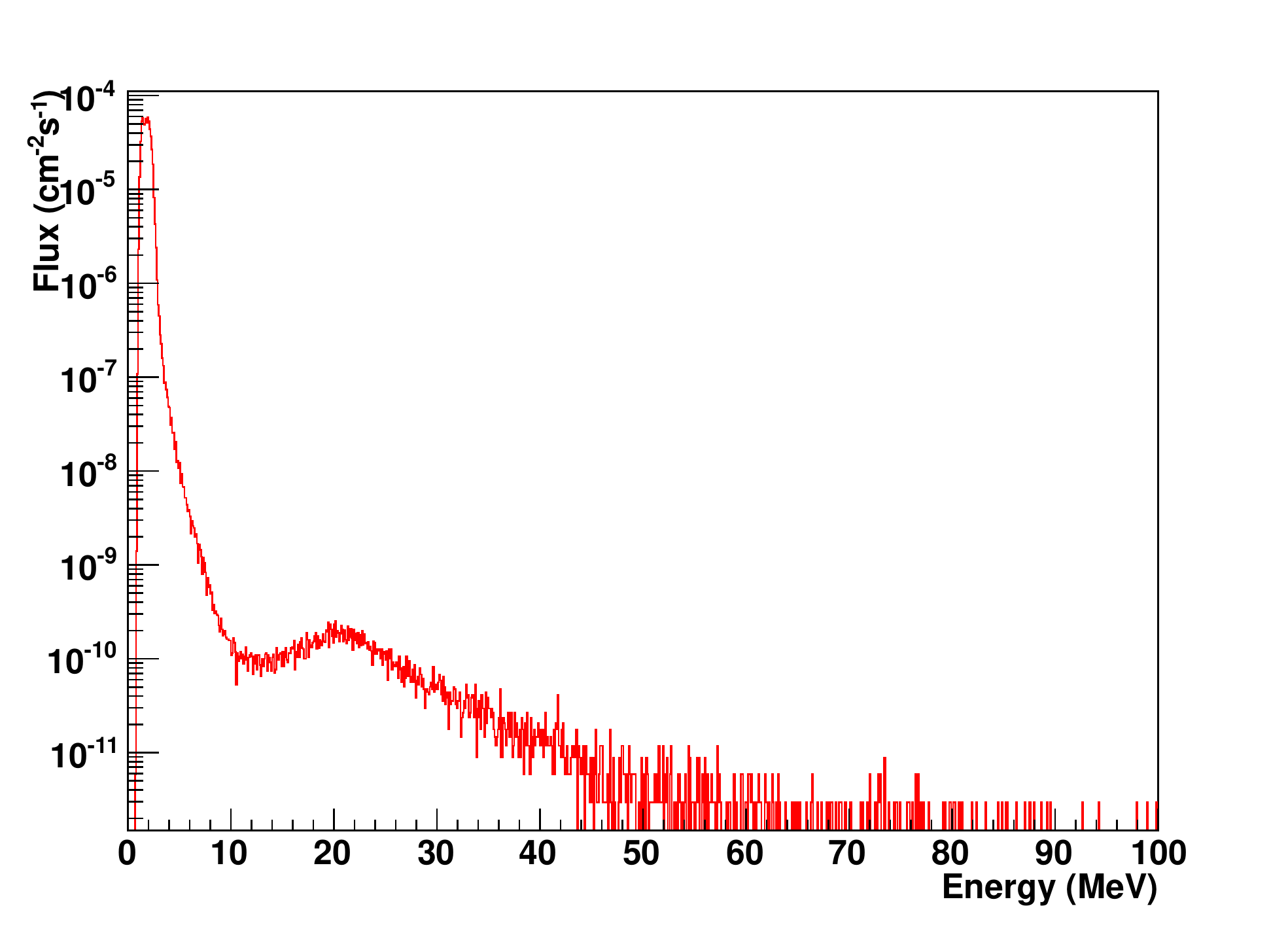}
\caption{\label{fullSpe}
The full energy spectrum in a 12 liters liquid scintillation detector with a live-time of 982.1 days.
}
\end{figure}

The experiment with a 12 liter liquid scintillation detector 
 was conducted at the Soudan Underground
Laboratory with a live-time of 982.1 days over 4 years. 
The detector is calibrated from $\sim$1 MeV to $\sim$20 MeV by using gamma ray sources
$^{22}$Na (1.275 MeV), AmBe (4.4 MeV), and the minimum ionization peak from
cosmic muons (20.4 MeV) \cite{czDet}. The energy response to the entire energy range is accumulated
and shown in Figure~\ref{fullSpe}. To maintain a stable energy scale over the entire experimental period, the
peak position of the muon minimum ionization is closely monitored. Energies are re-calibrated on weekly basis 
according to the variation of the peak position from the muon minimum ionization along with time. The pedestal value is also monitored and used in the correction of the pulse shapes when calculating energies. The $\gamma$ rays from radioactive decays ($^{40}$K, $^{232}Th$ and $^{238}$U) and ($\alpha$, $\gamma$) reactions induced by radon decays are also recorded and analyzed~\cite{chao}. The energies of $\gamma$ rays are significantly below 20 MeV. Muon-induced neutrons and ($\alpha$,n) neutrons have been reported in an earlier paper~\cite{czNeu} and the event rates are significantly smaller than the muon rate reported in this work.  
\par
Muons detected in our detector are largely suppressed by the overburden of the rock.
The variation of muon intensity is believed to correlated with the seasonal temperature variation
in the stratosphere
of atmosphere above the ground. An effective temperature $T_{eff}$ is defined using
a weighted average over the atmospheric depth\cite{minosfar}:
\begin{equation}\label{effT}
T_{eff} = \frac{\int_{0}^{\infty}\mathrm{d}XT(X)W(X)}{\int_{0}^{\infty}\mathrm{d}XW(X)}.
\end{equation}
Where $T(X)$ is atmospheric temperature in the stratosphere at a given atmospheric depth $X$, and the weight $W(X)$
is the temperature dependence of the production of mesons and their decay into muons
that can be observed in our detector.
The variation of atmospheric temperature in the stratosphere results in a change of the air density. 
Consequently, the change of the
air density modifies the ratio of meson decays to hadronic interaction and the hence
changing the muon flux observed underground. 
An effective temperature
coefficient $\alpha_{T}$ can be defined as:
\begin{equation}\label{alphaT}
\frac{\Delta I_{\mu}}{ < I_{\mu} > } = \alpha_{T}\frac{\Delta T_{eff}}{ < T_{eff} > }.
\end{equation} 

\par
\begin{figure}
\includegraphics[width=0.45\textwidth]{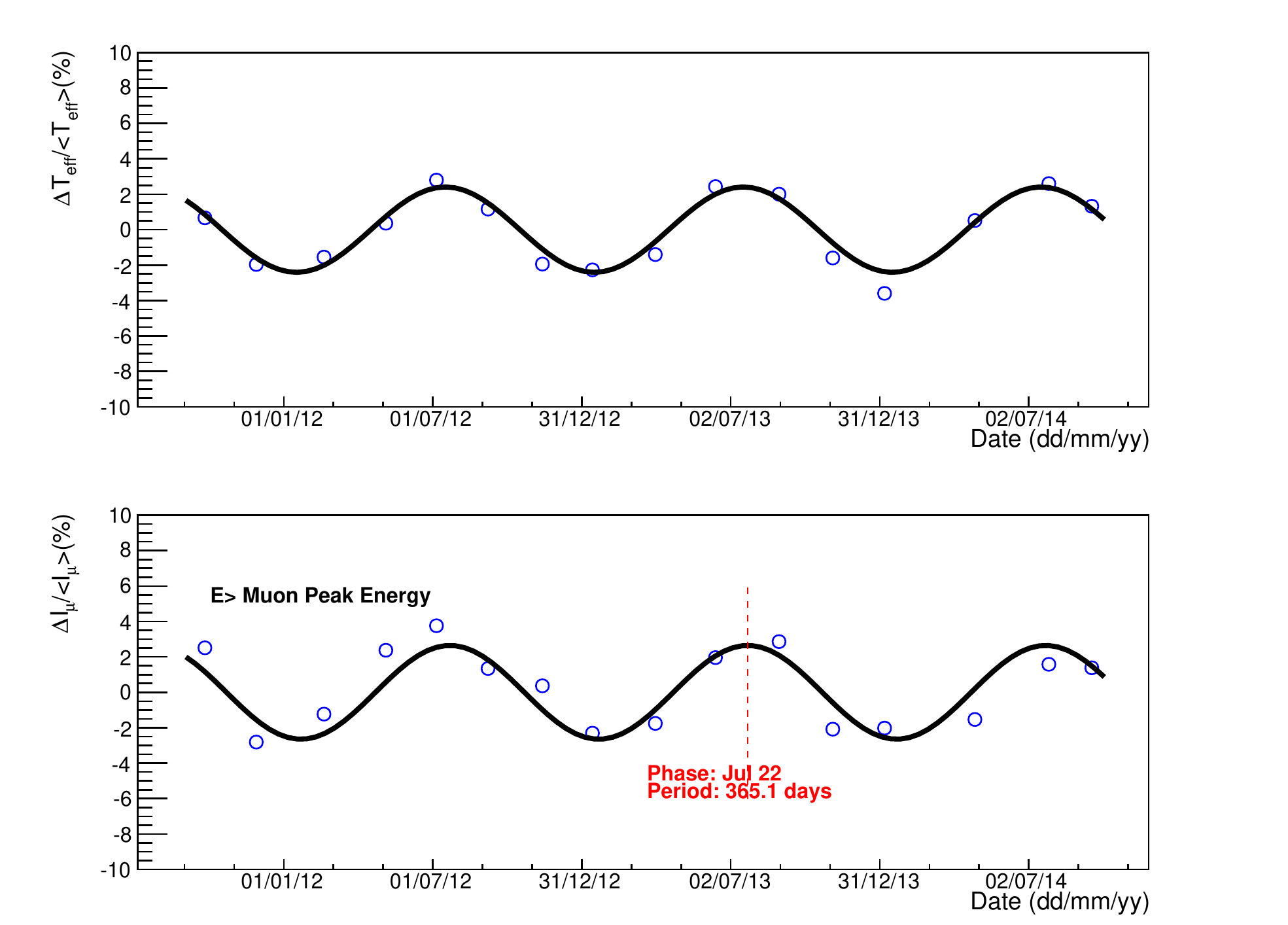}
\caption{\label{muongt20mev}
The top plot shows the effective temperature variation of the atmosphere above the Soudan Underground Laboratory. 
The data points in the bottom plot represent the variation of muon flux along time. The bin size is 64 days, corresponding to a total of 1836 muon events per bin. 
}
\end{figure}

Figure~\ref{muongt20mev} shows the variation of amplitude along with time.
The formula we use to determine the fractional modulation amplitude 
 $\delta I / \overline{ I }$ and the period $T$ is described
in Eq.~(\ref{percentRates}). 
\begin{equation}\label{percentRates}
I = \overline{I} + \Delta I = \overline{I} + \delta I \cos\left( \frac{2\pi}{T}(t-t_{0})\right).
\end{equation}
where $\overline{I}$ is the mean value and $\delta I $ is the variation amplitude. 
The phase $t_{0}$ is the time when the signal reaches its maximum. 
The top plot in Figure~\ref{muongt20mev} represents the effective temperature variation
of the atmosphere above the ground of the Soudan site. The atmospheric temperature data is
obtained from Ref.~\cite{uwyo}. A fixed period of 365.1 days is applied to fit the
variation pattern. The fitted variation amplitude is found to be 2.76\% with the phase at
Jul 12 $\pm$ 3.4 days. The bottom plot in Figure~\ref{muongt20mev} is the muon variation curve.
Data points with energy greater than the muon minimum ionization peak are collected to avoid 
the gamma ray contamination and any potential energy shift. With the fixed period of 365.1 days,
the fitted result gives the variation amplitude of 2.66\% with the phase at Jul 22 $\pm$ 36.2 days.

\par
\begin{figure}
\includegraphics[width=0.45\textwidth]{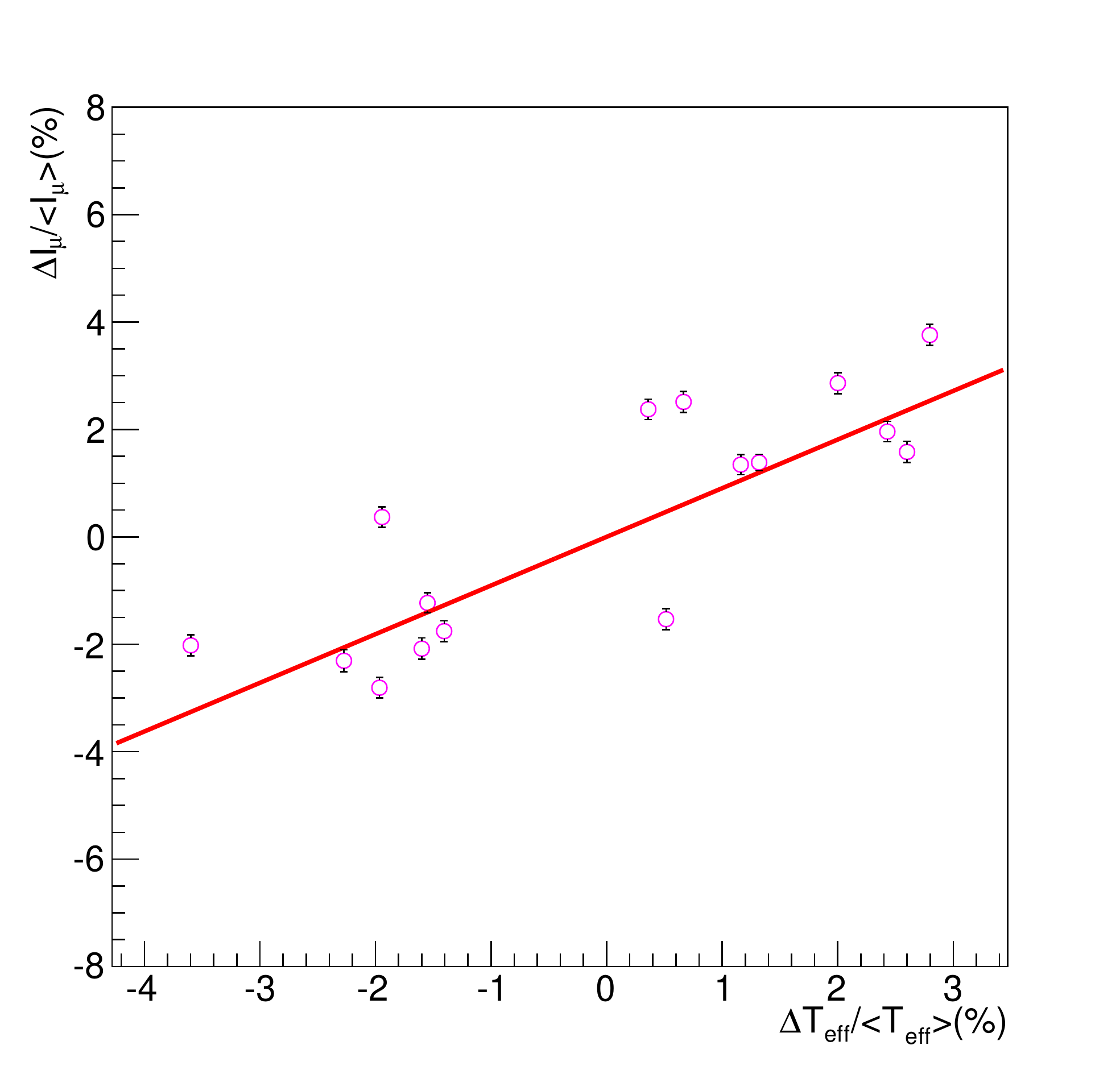}
\caption{\label{coeff}
The variation of muon rate $\Delta I_{\mu} / < I_{\mu} >$ as a function of $\Delta T_{eff}/ < T_{eff} >$. 
The slope of the linear fit gives $\alpha_{T} = 0.898 \pm 0.025$.
}
\end{figure}

The correlation of the percentage variation in the observed muon rate $\Delta I_{\mu} / < I_{\mu} > $ 
correlates with the change in effective temperature $ \Delta T_{eff} / < T_{eff} > $
is shown in Figure~\ref{coeff}. The fitting result determines the value of 
$\alpha_{T} = 0.898 \pm 0.025$. The error is dominated by the statistical uncertainty, since the systematical uncertainty is negligible. This is because the systematical uncertainty was carefully avoided using weekly calibration and the muon events were selected with energy greater than 20 MeV, which largely excludes gamma-ray ($E_{\gamma}$ $<$ 20 MeV) contamination. In addition, the event rate from the muon-induced neutrons is much smaller than the muon event rate detected in the detector. 
The remaining uncertainty associated with the value of $\alpha_{T}$ measured in this work is therefore governed by the
limited statistical error per bin (64 days with 1836 muon events per bin).
\par
\begin{figure}
\includegraphics[width=0.45\textwidth]{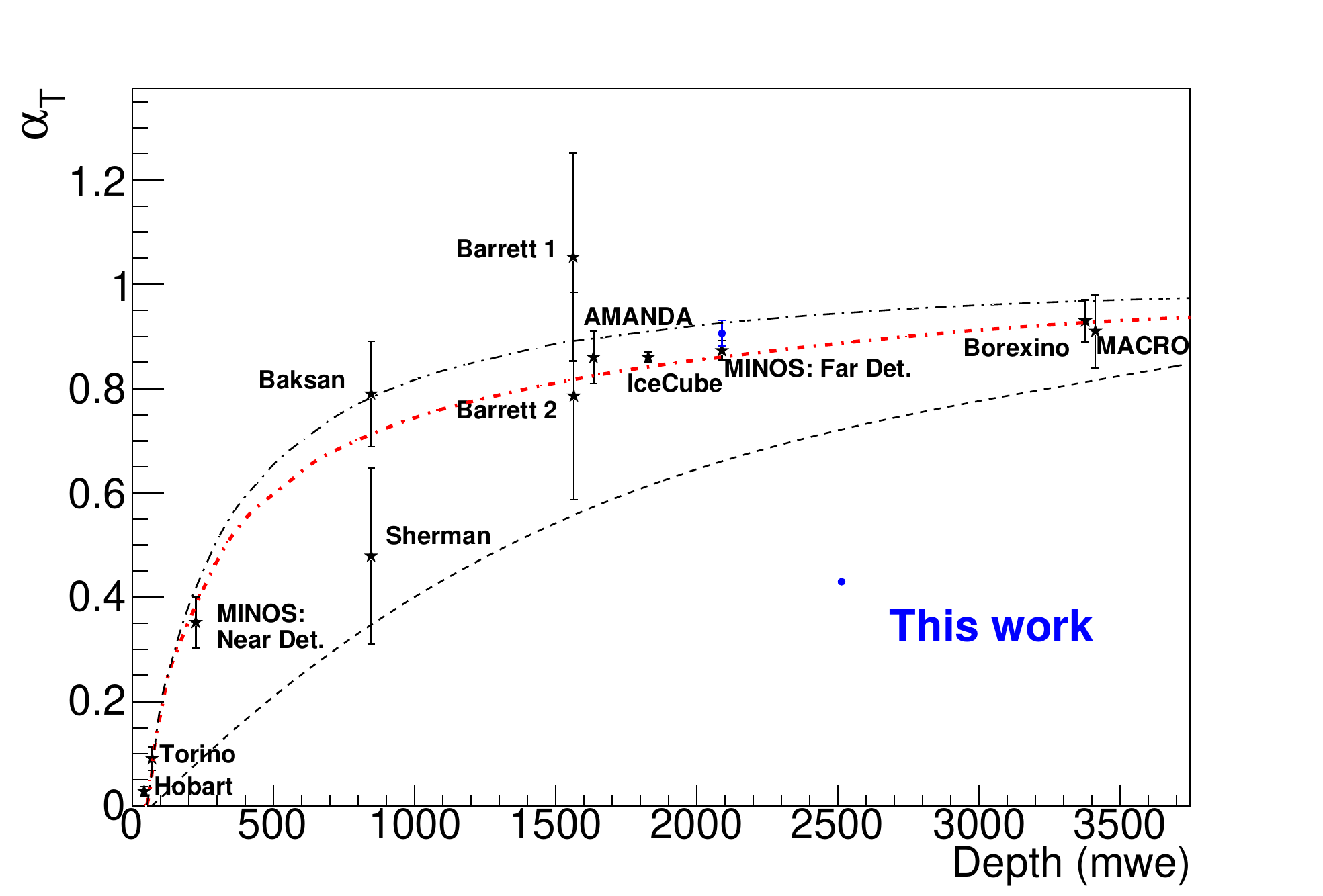}
\caption{\label{expComp}
Shown is the values of $\alpha_{T}$ at various detector depth. The experimental results are labelled by the name of collaboration. 
The red line is the value predicted including muon production by pions and kaons. 
The dashed curves above and below stand for pions or kaons only, respectively. 
}
\end{figure}

Figure~\ref{expComp} summarizes the measured values for $\alpha_{T}$ from various underground depths. 
The reported values at different underground sites agree with the predicted $\alpha_{T}$ 
(red curve in Figure~\ref{expComp}) well. 
Our detector is adjacent to the MINOS far detector at the same depth level of the Soudan Underground Laboratory.
Both results show a good agreement with the prediction.   
 
The charged $K$/$\pi$ ratio, $r(K/\pi$), can be determined using the relation below~\cite{gra}:
\begin{equation}
    r(K/\pi) = \frac{(\alpha_T)_\pi/\alpha_T - 1} { 1- (\alpha_T)_{K}/\alpha_T},
    \label{a1}
\end{equation}
where $(\alpha_{T})_{K,\pi}$ can be obtained using the theoretical prediction~\cite{gra}:
\begin{equation}
    (\alpha_T)_{K,\pi} = 1/[\frac{\gamma}{\gamma+1} \cdot \frac{\epsilon_{K,\pi}}{1.1E_{th}cos\theta} + 1].
    \label{a2}
\end{equation}
Utilizing the muon spectrum index, $\gamma$ =1.7$\pm$0.1, kaon critical energy $\epsilon_{k}$ = 0.851$\pm$0.014 TeV, and pion critical energy $\epsilon_{\pi}$ = 0.114$\pm$0.003 TeV given by the Particle Data Group~\cite{pdg} and E$_{th}$cos$\theta$ = 0.795$\pm$0.14 TeV from MINOS~\cite{minosfar}, we can obtain $(\alpha_T)_{K}$ = 0.620$^{+0.029}_{-0.037}$ and $(\alpha_{T})_{\pi}$ = 0.924$^{+0.008}_{-0.011}$. Plugging the values of $(\alpha_{T})_{K,\pi}$ and the measured $\alpha_{T}$ = 0.898$\pm$0.025 into equation~\ref{a1}, we obtain $r(K/\pi)$ = 0.094$^{+0.044}_{-0.061}$. This is consistent with $r(K/\pi$) = 0.12$^{+0.07}_{-0.05}$ determined by MINOS~\cite{minosfar}. To further constrain the uncertainty of $r(K/\pi$) from the measurements, we conduct a Geant4 simulation to study the correlation between the temperature variation in the stratosphere and the muon rate annual modulation underground at the Soudan Underground Laboratory for a given $r(K/\pi$). 

\section{Simulation of the muon rate annual modulation with a given $K$/$\pi$ ratio}
\subsection{Simulation of muons from primary cosmic-ray Protons }
As observed in Figure~\ref{muongt20mev}, the muon rate modulates over a year period. 
The correlation with the variation of temperature is demonstrated in Figure~\ref{coeff}.
In the summer time, the temperature in the stratosphere
increases. As a result, the air density decreases. Therefore, more mesons undergo decay processes to 
produce more energetic muons, which can be observed in a detector underground. In contrast to the summer,
the temperature in the stratosphere decreases in the winter time, which increases the air density. Thus, 
more mesons can interact with air particles to produce muons with lower energies, which have less chance to reach a detector deep underground. This phenomenon is observed as the muon rate annual modulation 
in a detector underground. 
Since kaons (K$^{+}$ and K$^{-}$ have a shorter half-life (12.4 $ns$) than pions 
(26 ns for $\pi^{+}$ and $\pi^{-}$)~\cite{pdg}, 
it is expected that this phenomenon is mainly due to the change of the fraction of pions that undergo decays
with respect to the interactions with air particles~\cite{gra} in the stratosphere. Accordingly, the 
correlation between the observed muon rate annual modulation and the variation of temperature in the stratosphere
is sensitive to the $K$/$\pi$ ratio in the production place. As a result, measuring this correlation provides
an indirect way to measure the $K$/$\pi$ ratio induced by very high energy cosmic rays in the stratosphere.

The atmospheric $K$/$\pi$ ratio was first measured using the MINOS-FD data in 2009~\cite{minosfar}. The measured
$\alpha_{T}$, 0.873$\pm$0.009(stat)$\pm$0.010(syst) was used to determine the $K$/$\pi$ ratio together with
the theoretical prediction with large errors up to 40\%~\cite{gdb}. The determined 
atmospheric $K$/$\pi$ ratio is 0.12$^{+0.07}_{-0.05}$. Utilizing a well-understood 12-liter liquid scintillation detector, we have determined $\alpha_T$ = 0.898$\pm$0.025, which corresponds to a $K/\pi$ ratio of 0.094$^{+0.044}_{-0.061}$. To further constrain the uncertainty on the value of $K/\pi$ ratio, 
a simulation is performed to reproduce the surface muons originating from primary cosmic rays with a given $K/\pi$ ratio to examine the correlation between the 
muon rate annual modulation and the variation of temperature in the stratosphere. 

\begin{figure}
\includegraphics[width=0.45\textwidth]{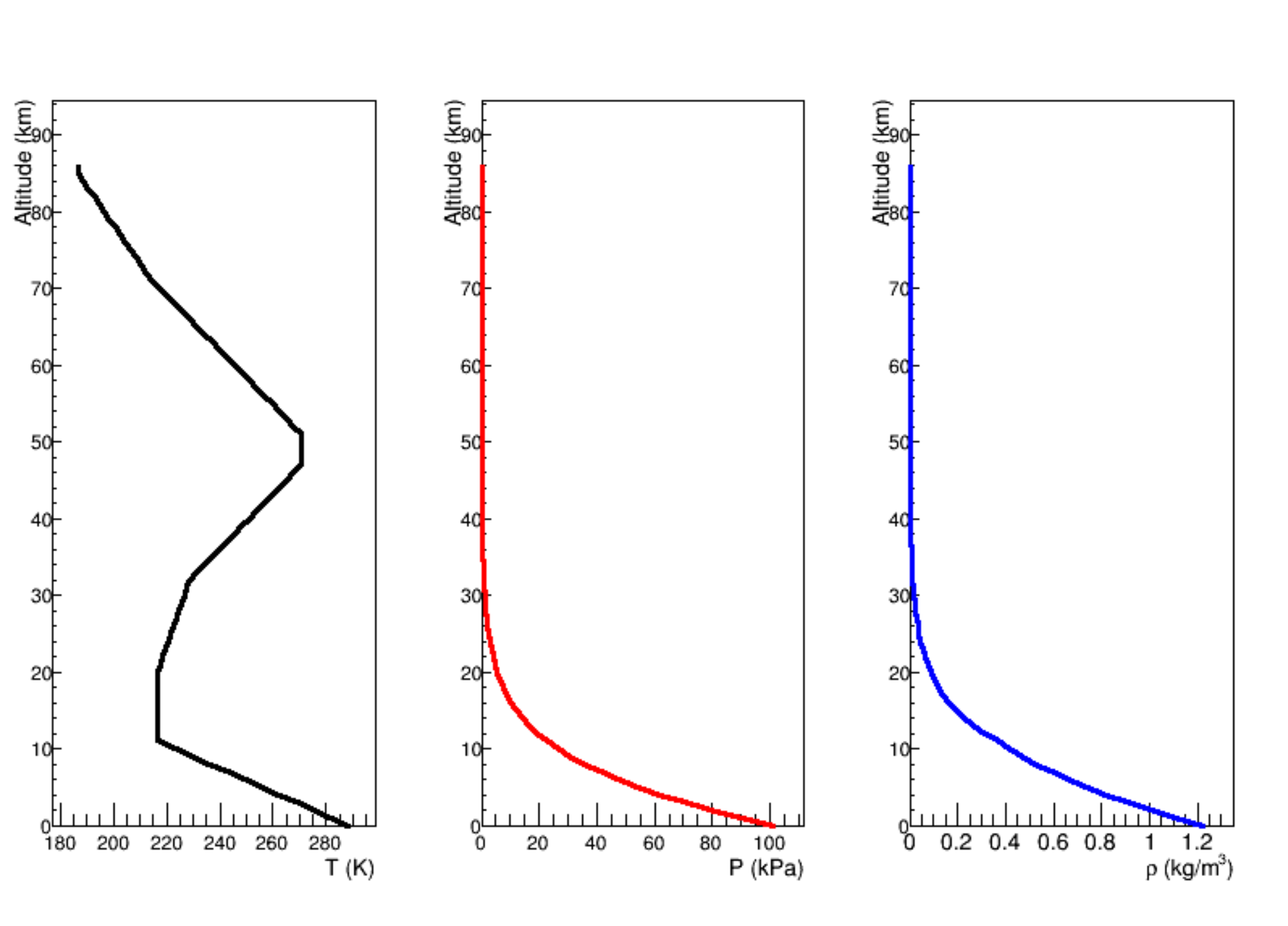}
\caption{\label{atmospheric}
Shown is the U.S. 1976 Atmospheric Model~\cite{airModel} with 100 layers evenly divided for altitude ranging from 0 to 100 km.
}
\end{figure}

\begin{figure}
\includegraphics[width=0.45\textwidth]{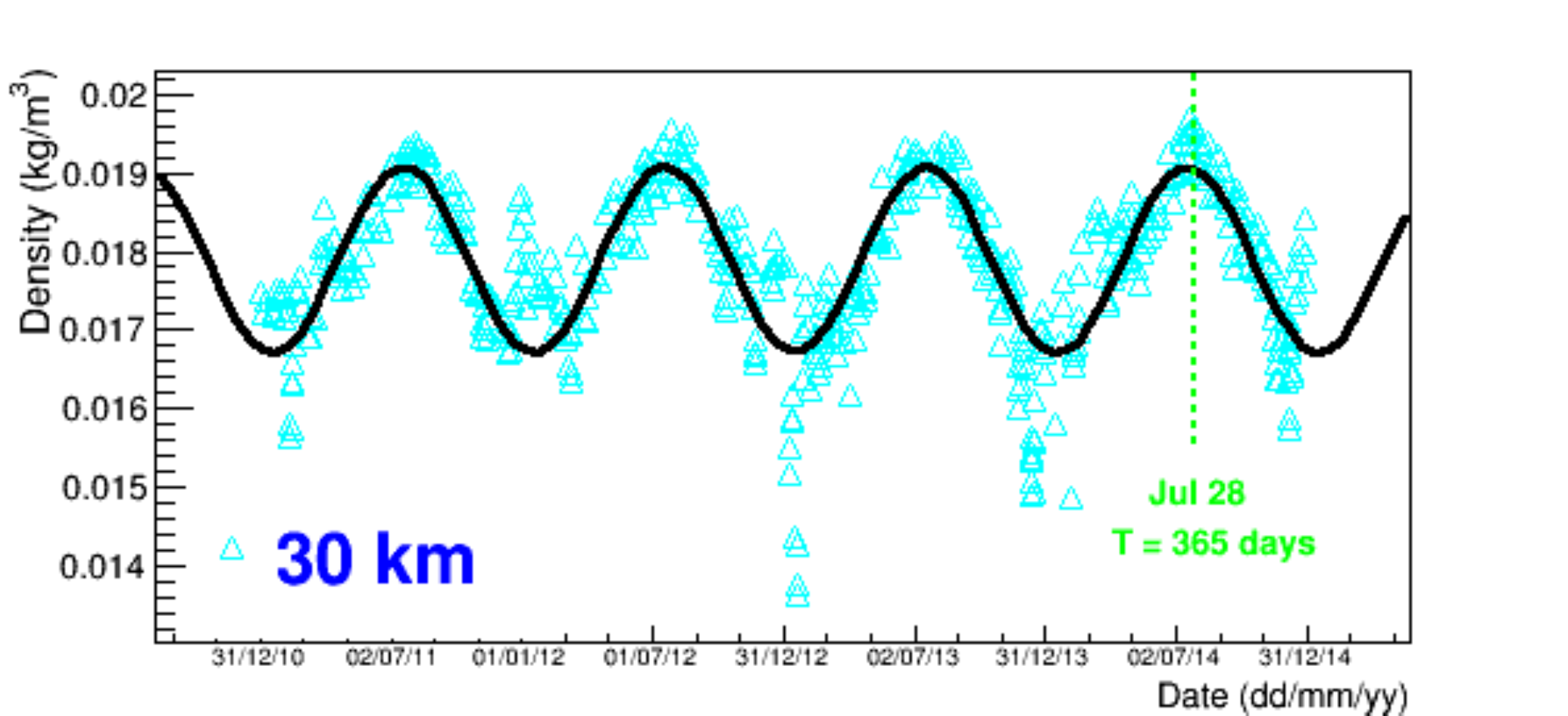}
\caption{\label{igra}
Shown is the seasonal air density variation obtained from IGRA for the location close to Soudan.
}
\end{figure}

Primary cosmic-ray protons are generated at the top surface 100 km above sea level, since high-energy particles arriving from outer space are mainly (89\%) protons~\cite{pricos}. 
We cast the proton energy range from 1 GeV to 100 TeV with the differential spectral index to be -2.7 ($\alpha$ = $\gamma$ +1)~\cite{pdg}.   
The U.S. 1976 Atmosphere Model~\cite{airModel} has been adopted to simulate the average air density and pressure 
change along with the altitude as shown in Figure~\ref{atmospheric}. 
The seasonal air density variation from the Integrated Global Radiosonde Archive (IGRA) for the location 
which is very close to Soudan~\cite{igra} is adopted in the simulation. As an example, we show the
seasonal air density variation in the stratosphere at the level of 30 km above sea level in Figure~\ref{igra}.
The 
Geant4 module physics QGSP\_BERT\_EMV~\cite{geant4} with the step length 10 cm is applied
in the simulation to reveal small perturbations of atmospheric weights caused by the variation of temperature. The hadronic models used in Geant4 are Bertini cascade (0 - 5.0 GeV) and 
Fritiof with Precompound (FTFP) (4.0 GeV - 100 TeV) for hadrons as well as G4GammaNuclearReaction (0 - 3.5 GeV)
and Quark-gluon String with Precompound (QGSP) (3.0 GeV - 100 TeV) for any secondary high-energy gamma rays~\cite{geant44}. 

\subsection{Simulation results}
\begin{figure}
\includegraphics[width=0.45\textwidth]{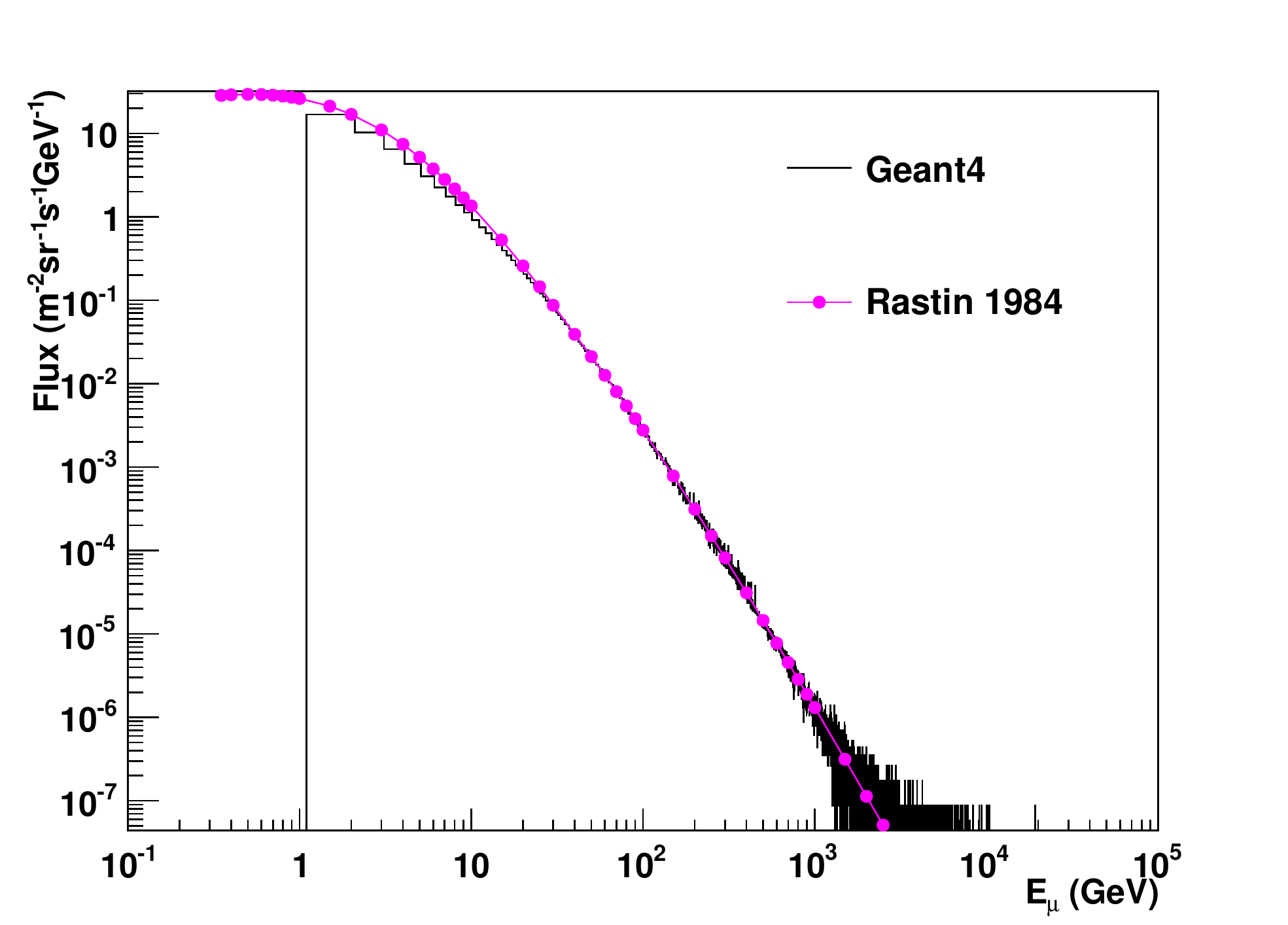}
\caption{\label{muonSpectraComp}
Simulated muon energy spectrum at sea level compared with the measurement~\cite{rastin}. 
}
\end{figure}

Figure~\ref{muonSpectraComp} shows the simulated cosmic-ray muon energy spectrum in comparison with the measurement~\cite{rastin}. 
Cosmic muons are secondary particles from cosmic ray air shower events. Muons reaching the sea level are
collected and the simulated energy spectrum is compared with the experimental data (see FIG.~\ref{muonSpectraComp}).  
\par
\begin{figure}
\includegraphics[width=0.45\textwidth]{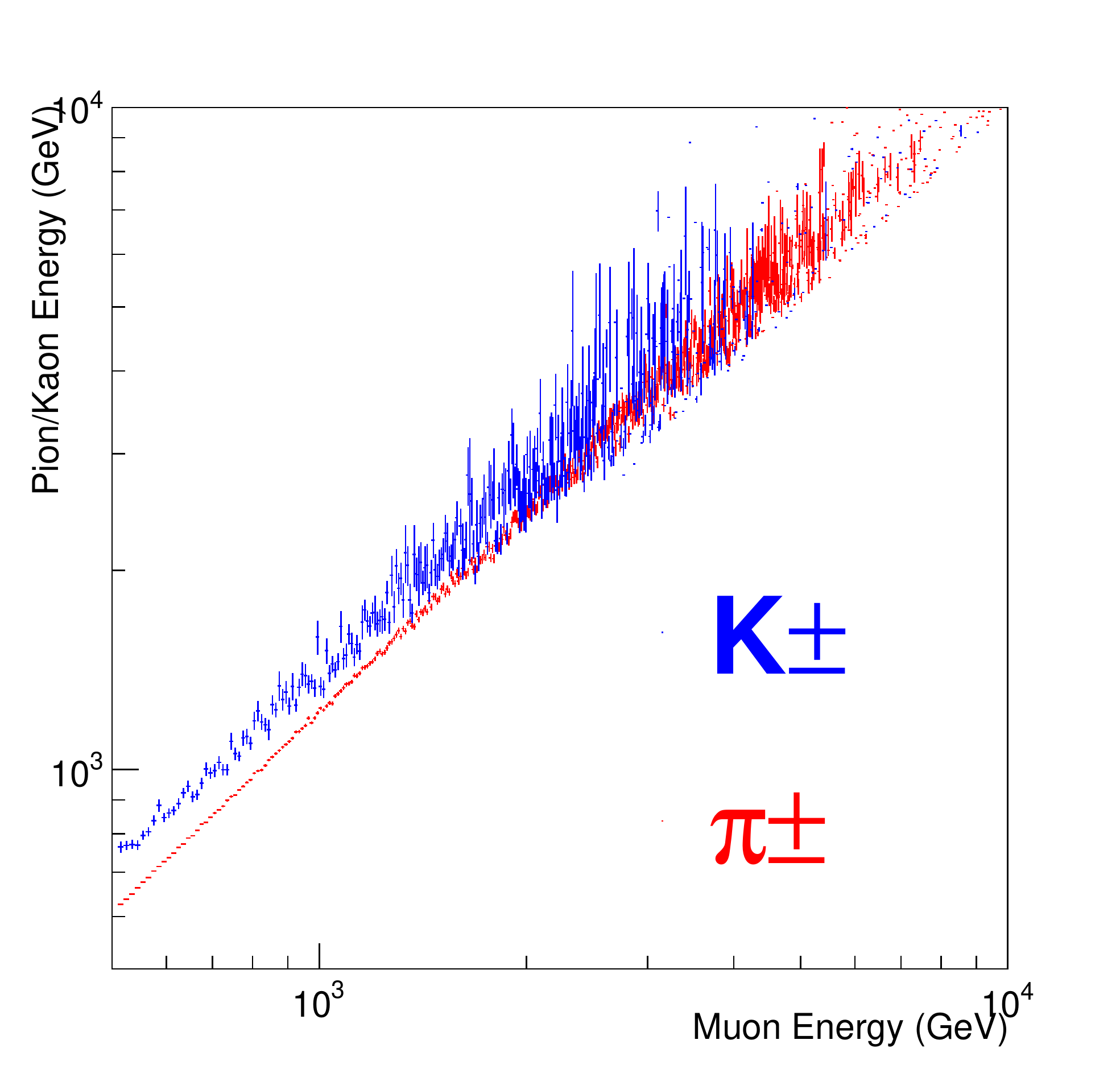}
\caption{\label{parentEn}
The average energy conversion of parent $\pi\pm, k\pm$ versus secondary muons.
}
\end{figure}

For our experimental setup, only those surface muons with energy greater than 500 GeV can reach the underground depth
where our detector is located. The parent particle of those muons are mainly cosmic ray $k\pm$ and $\pi\pm$. The charged $K$/$\pi$ 
with their energy greater than 0.5 TeV is counted as: 0.1598.
The average energy conversion between the parent $k\pm,\pi\pm$ and secondary muons are shown in FIG.~\ref{parentEn}. 
Most of decays of $k\pm$ and $\pi\pm$ are found to be around 20$\sim$30 km above the sea level. 

\begin{figure}
\includegraphics[width=0.45\textwidth]{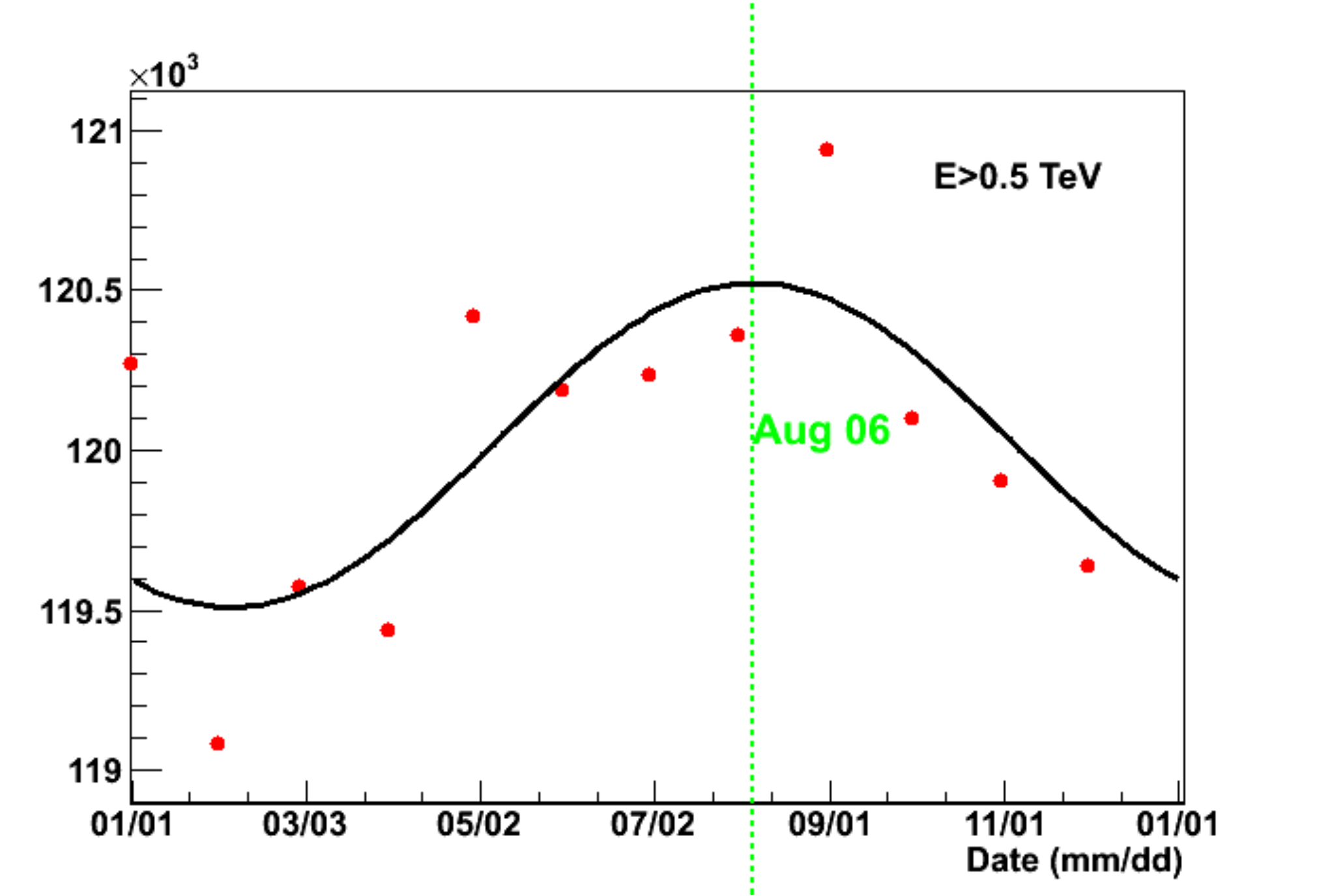}
\caption{\label{muon500GeV}
The expected muon rate annual modulation for muon energy greater than 0.5 TeV. The modulation amplitude is only 0.42\%, which is much less than the experimental measurement of 2.66\%. 
}
\end{figure}

\begin{figure}
\includegraphics[width=0.45\textwidth]{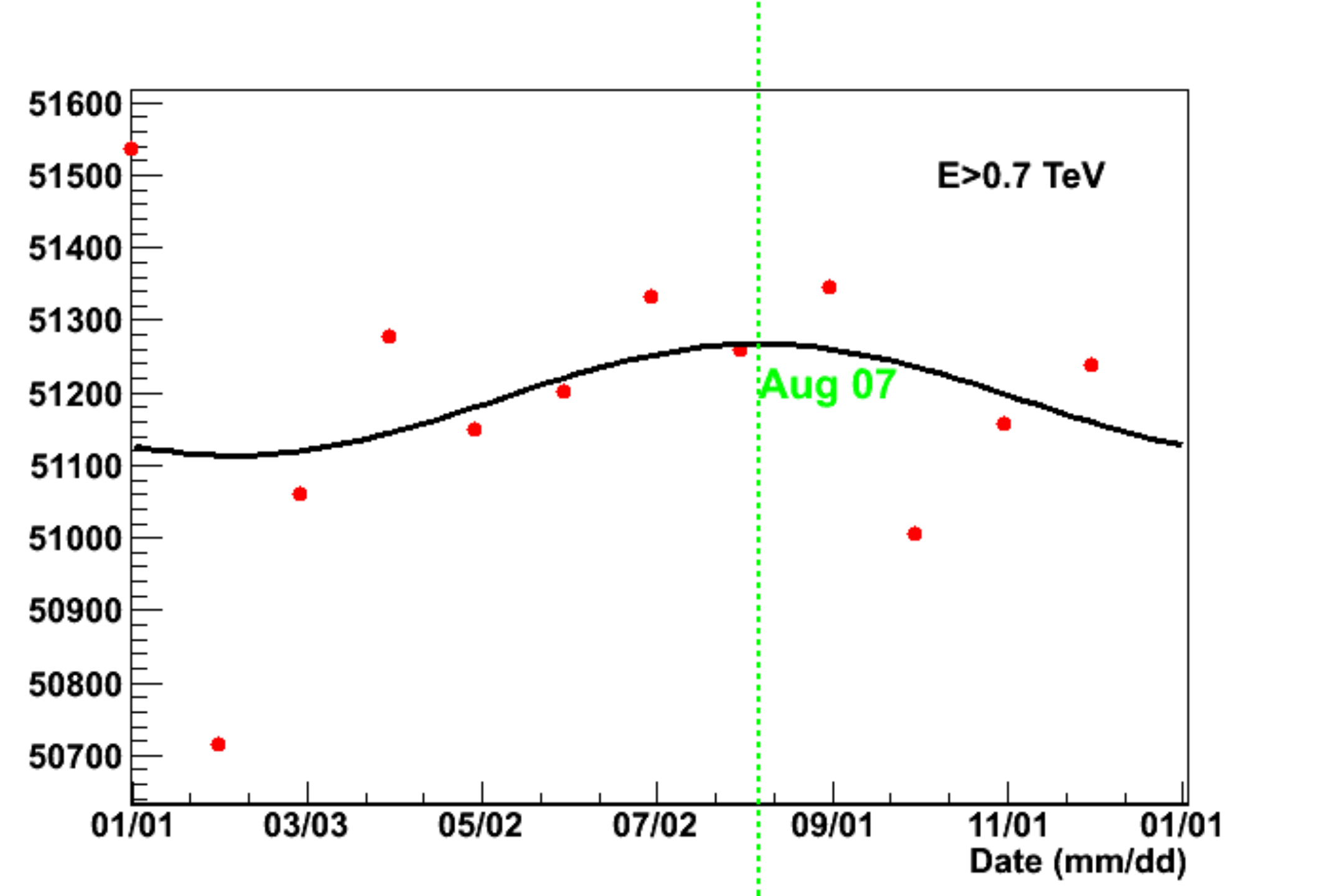}
\caption{\label{muon700GeV}
The expected muon rate annual modulation for muon energy greater than 0.7 TeV. The modulation amplitude is only 0.15\%, which is much less than the experimental measurement of 2.66\%. 
}
\end{figure}

With a charged $K$/$\pi$ ratio of 0.1598 in the simulation and the seasonal air density variation shown in Figure~\ref{atmospheric},
 the muon rate annual modulation is not observed in the simulation for muons with energies up to 0.5 TeV and 0.7 TeV 
as shown in Figure~\ref{muon500GeV} and Figure~\ref{muon700GeV}.
 The results from the simulation indicate that the charged $K$/$\pi$ ratio of 0.1598 originated from cosmic-ray protons interacting with air particles cannot generate the sufficient
muon rate annual modulation observed in a detector at the Soudan Underground Laboratory. 
Since this $K/\pi$ ratio given by the Geant4 simulation is greater than the upper bound of 0.138 determined by this work. This simulation sheds light on the constrain of the charged $K/\pi$ ratio, which is in favor of the measured value 0.0924$^{+0.044}_{-0.061}$. This lower $K/\pi$ ratio is also consistent with the hypothesis of the muon rate annual modulation is mainly due to the change of the fraction of pions that undergo decays
with respect to the interactions with air particles~\cite{gra} in the stratosphere. A higher $K/\pi$ ratio would reduce the muon rate annual modulation because kaons are not as sensitive as pions to the temperature variation in the stratosphere due to a relatively short half-life.  
Note that the Soudan Underground Laboratory has a flat surface. We used the flat Earth approximation to simulation muons traversing the overburden with an average density of 2.85 g/cm$^3$~\cite{chao}. 

\section{Conclusions}
\begin{figure}
\includegraphics[width=0.45\textwidth]{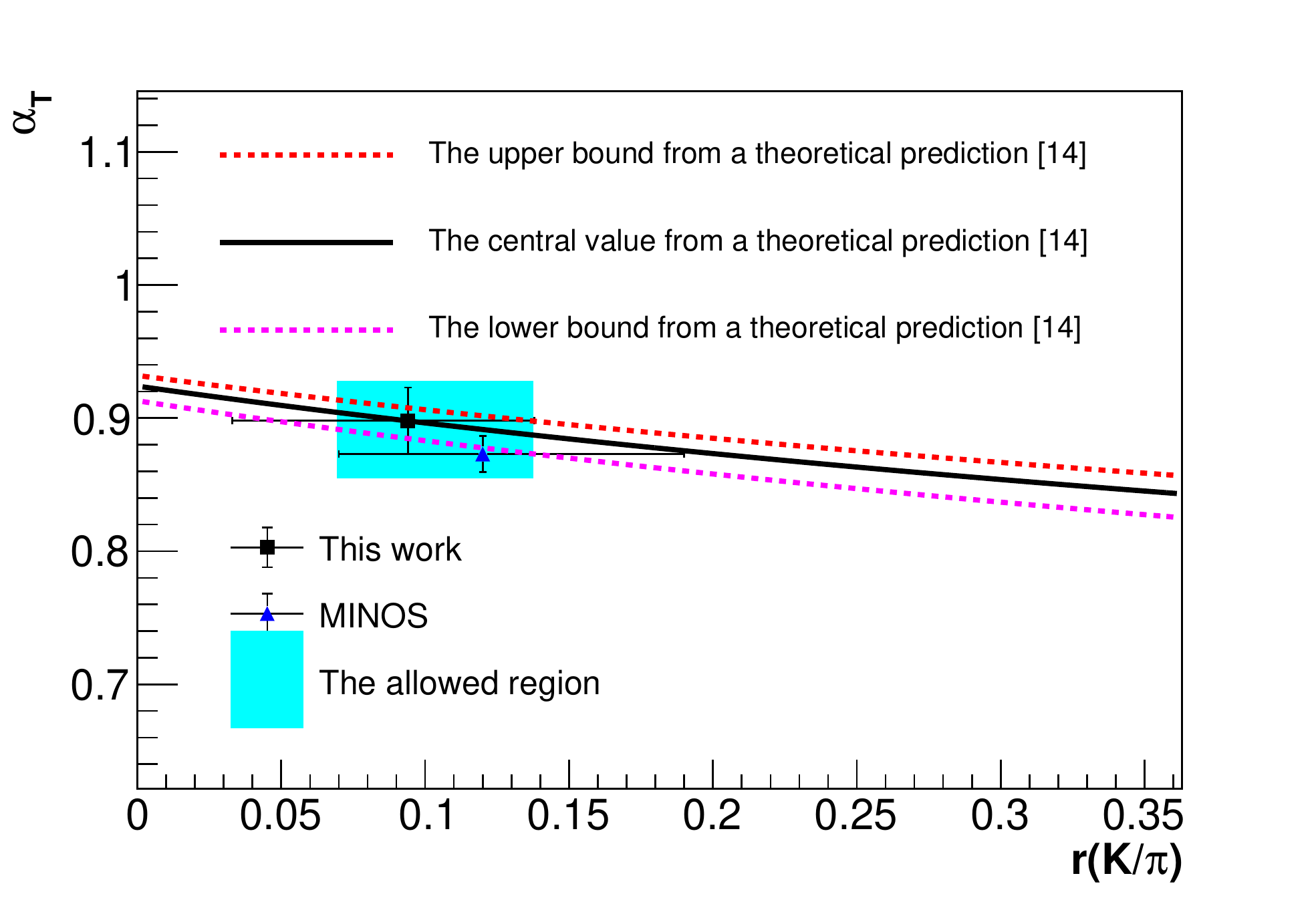}
\caption{\label{constrain}
Shown is $\alpha_{T}$ versus the charged $K/\pi$ ratio, $r(K/\pi)$. The allowed region is determined by combining the measurement from MINOS with the measurement from this work. 
}
\end{figure}

Our detector accumulated data at the Soudan Underground Laboratory (2100 m.w.e.) for over four years. 
Data analysis gives a muon flux $I_{\mu} = (1.65\pm0.12)\times 10^{-7} cm^{-2}s^{-1}$. 
Seasonal modulation of muon rates are observed with the percentage amplitude of 2.66\%
and the phase to be Jul 22 $\pm$ 36.2 days. The correlation between atmospheric temperature
variations and the changes in the muon rates observed in our detector has been investigated. 
The temperature coefficient of $\alpha_{T} = 0.898 \pm 0.025$ is found for the underground depth
where our detector is located. This result is in a good agreement with the measurement made by the MINOS-FD (0.873$\pm$0.009(stat.)$\pm$0.010(syst.)). The 
value of $\alpha_{T}$, 0.898$\pm$0.025, implies that the atmospheric $K$/$\pi$ ratio is 0.094$^{+0.044}_{-0.061}$ in the stratosphere determined by this work. Utilizing the Geant4 simulation, we find out a charged $K$/$\pi$ ratio of 0.1598, greater than 0.138 the upper bound of this work, cannot contribute to the observed the muon annual modulation 
in our detector at the Soudan Underground Laboratory. If one combines this work with the measurement of MINOS (0.12$^{+0.07}_{-0.05}$), the charged $K$/$\pi$ ratio is constrained to a range between 0.07 to 0.0138, as shwon in Figure~\ref{constrain} determined by the observed muon annual modulation at the depth of Soudan.   

\section*{Acknowledgments}
The authors wish to thank Fred Gray, Keenan Thomas,
Anthony Villano, Priscilla Cushman and the Soudan underground facility management for their invaluable
suggestions and help. We also would like to thank Christina Keller for a careful reading of this manuscript and Jing Liu and Arun Soma for their useful discussion.  This work was supported in part by NSF PHY-0919278, NSF PHY-1242640, NSF OISE 1743790, DOE grant DE-FG02-10ER46709, the Office of Research at the University of South Dakota and a research center supported by the State of South Dakota.

\end{document}